\documentclass[12pt]{article}

\usepackage{dina4pcg}
\usepackage{epsfig}
\usepackage{amssymb}

\def\oo{\'o}

\def\mz{M_Z}
\def\asz{\as(\mz)}
\def\as{\alpha_s}
\def\z0{Z^0}
\def\qq{q\bar q}
\def\q2{Q^2}
\def\eprr{ep\rightarrow e\ +\ {\rm jet}\ +\ {\rm jet}\ + \ {\rm X}}
\def\etjet{E_T^{\rm jet}}
\def\etajet{\eta^{\rm jet}}
\def\kt{k_T}
\def\g2{GeV$^2$}
\def\etlab{E_{T,{\rm LAB}}^{\rm jet}}
\def\etalab{\eta^{\rm jet}_{\rm LAB}}
\def\etjb{E^{\rm jet}_{T,{\rm B}}}
\def\asmz#1#2#3#4#5#6{\asz = #1\pm #2\ {\rm (stat.)}\ ^{+#4}_{-#3}\ {\rm (exp.)}\ ^{+#6}_{-#5}\ {\rm (th.)}}
\def\etaphi{\eta-\varphi}
\def\etar{-1<\etajet<2.5}
\def\wrn{$142<\wgp<293$ GeV}
\def\wgp{W_{\gp}}
\def\gp{\gamma p}
\def\ccos{\cos\theta^*}
\def\seta{d\sigma/d\etajet}
\def\cost{\vert\cos\theta^*\vert}
\def\sccos{d\sigma/d\ccos}
\def\colab#1{#1 Coll.}
\def\etal{et al.}
\def\scost{d\sigma/d\cost}
\def\Journal#1#2#3#4{{#1} {#2} (#3) #4}

\def\NPB{{\em Nucl. Phys.} {\bf B}}
\def\PLB{{\em Phys. Lett.}  {\bf B}}
\def\PRL{{\em Phys. Rev. Lett.}}
\def\PRD{{\em Phys. Rev.} {\bf D}}

\def\ZPC{{\em Z. Phys.} {\bf C}}

\def\EPC{{\em Eur. Phys. Jour.} {\bf C}}

\def\CPC{{\em Comp. Phys. Comm.}}

\def\SJN{{\em Sov. J. Nucl. Phys.}}
\def\SJJ{{\em Sov. Phys. JETP}}

\begin{document}

\title{\bf Jet production at HERA\footnote{Talk given at the ``XXXIV
    International Symposium on Multiparticle Dynamics'', Sonoma
    County, California, USA, July $26^{\rm th}$ - August $1^{\rm st}$,
    2004.}}

\author{C. Glasman\thanks{Ram\oo n y Cajal Fellow.}\\
(on behalf of the ZEUS and H1 Collaborations)\\
Universidad Aut\oo noma de Madrid, Spain}

\date{}

\maketitle

\begin{abstract}
Recent results from jet production in deep inelastic $ep$ scattering
to investigate parton dynamics at low $x$ are reviewed. The results on
jet production in deep inelastic scattering and photoproduction used
to test perturbative QCD are discussed and the values of $\asz$
extracted from a QCD analysis of the data are presented.
\end{abstract}

\section{Introduction}
Jet production in neutral-current (NC) deep inelastic $ep$ scattering
(DIS) and photoproduction provide tests of perturbative QCD (pQCD)
calculations and of the parametrisations of the parton distribution
functions (PDFs) in the proton. Jet cross sections allow the
determination of the fundamental parameter of QCD, the strong coupling
constant $\as$, and help to constrain the proton PDFs.

\begin{figure}[ht]
\setlength{\unitlength}{1.0cm}
\begin{picture} (10.0,10.0)
\put (3.0,5.0){\epsfig{figure=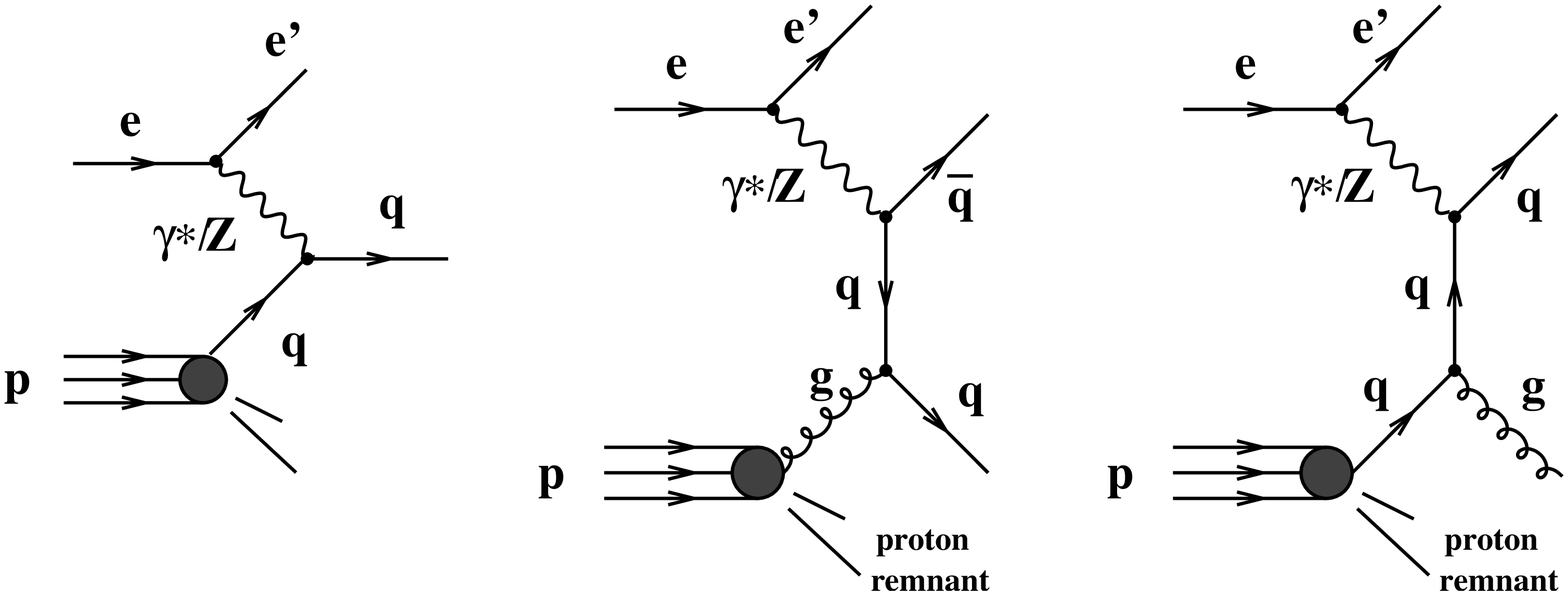,width=12.0cm}}
\put (3.5,0.0){\epsfig{figure=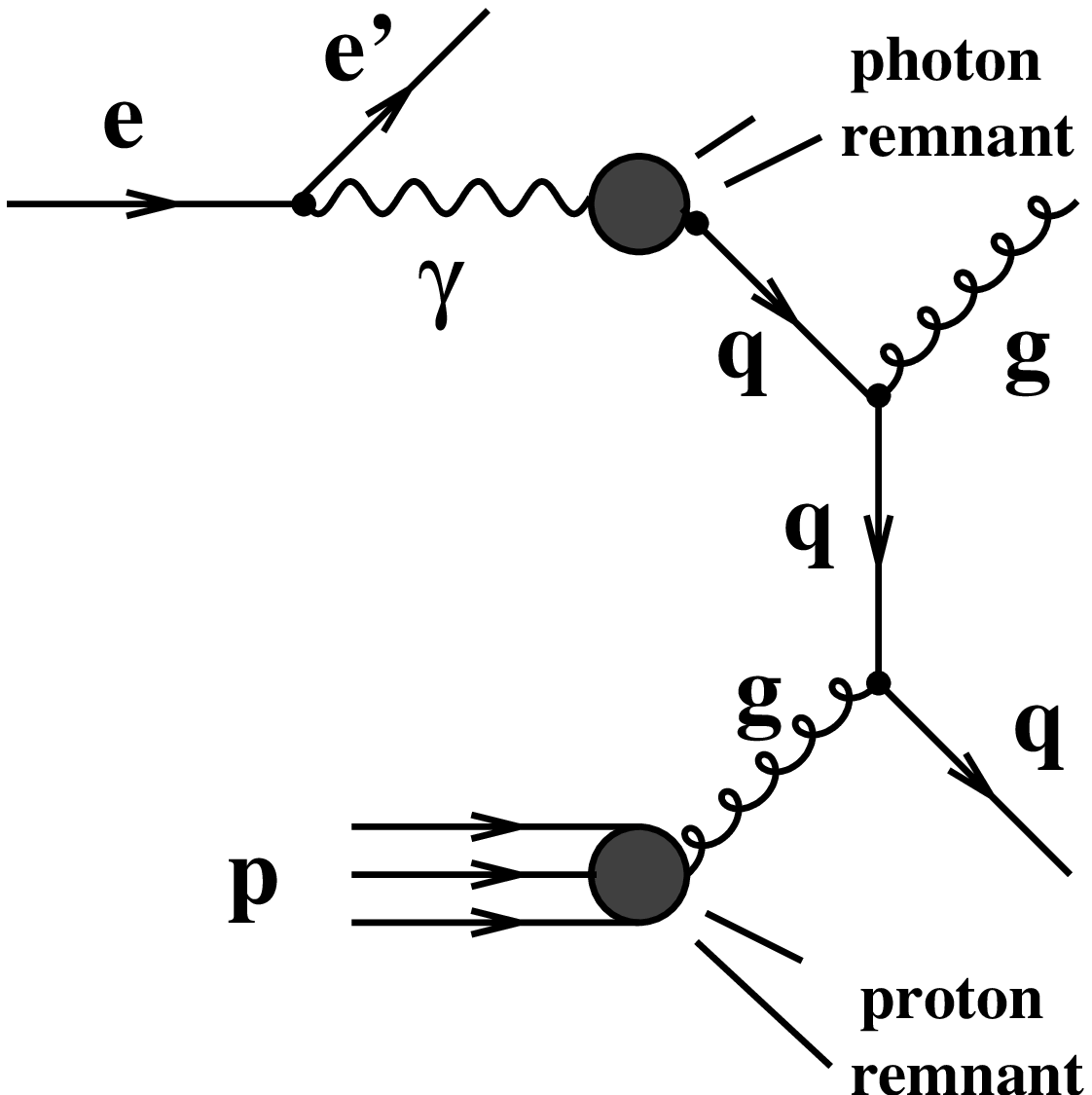,width=4.0cm}}
\put (10.0,0.0){\epsfig{figure=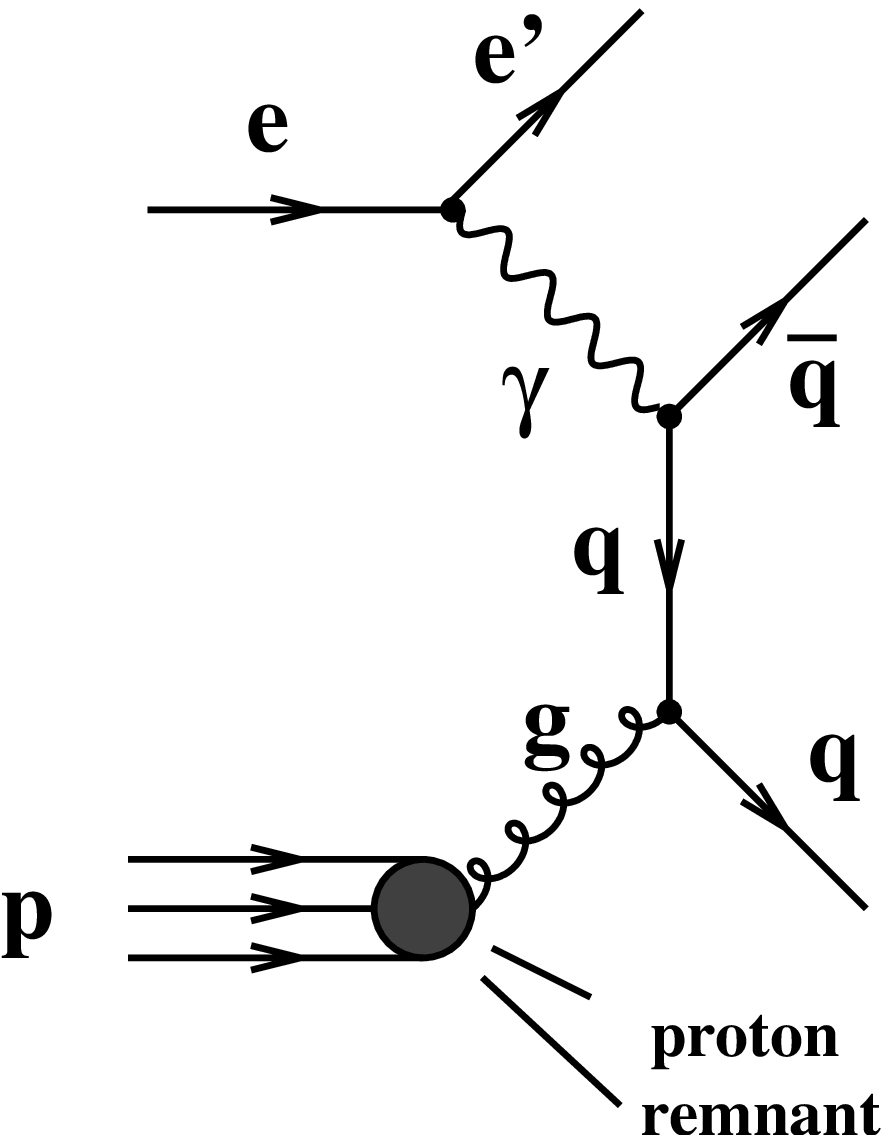,width=3.5cm}}
\put (4.2,5.7){\bf\small (a)}
\put (8.5,5.0){\bf\small (b)}
\put (12.9,5.0){\bf\small (c)}
\put (5.5,0.0){\bf\small (d)}
\put (11.4,0.0){\bf\small (e)}
\end{picture}
\caption{Examples of Feynman diagrams for deep inelastic $ep$
  scattering processes: (a) quark-parton model, (b) boson-gluon fusion
  and (c) QCD Compton. Examples for photoproduction processes: (d)
  resolved and (e) direct.
  \label{one}}
\end{figure}

Up to leading order (LO) in $\as$, jet production in NC DIS
proceeds via the quark-parton model (QPM) ($Vq\rightarrow q$, where
$V=\gamma^*$ or $\z0$, Fig.~\ref{one}a), boson-gluon fusion (BGF)
($Vg\rightarrow \qq$, Fig.~\ref{one}b) and QCD-Compton (QCDC)
($Vq\rightarrow qg$, Fig.~\ref{one}c) processes. The jet production
cross section is given in pQCD by the convolution of the proton PDFs
and the subprocess cross section,
$$d\sigma_{\rm jet}=\sum_{a=q,\bar q,g}\int dx\ f_a(x,\mu_{F_p})\ d\hat \sigma_a(x,\as(\mu_R),\mu_R,\mu_{F_p}),$$
where $x$ is the fraction of the proton's momentum taken by the
interacting parton, $f_a$ are the proton PDFs, $\mu_{F_p}$ is the
proton factorisation scale, $\hat\sigma_a$ is the subprocess cross
section and $\mu_R$ is the renormalisation scale.

The main source of jets at HERA is hard scattering in photon-proton
(photoproduction) interactions in which a quasi-real photon
($\q2\approx 0$, where $\q2$ is the virtuality of the photon) emitted
by the electron beam interacts with a parton from the proton to
produce two jets in the final state. In LO QCD, there are two
processes which contribute to the jet photoproduction cross section:
the resolved process (Fig.~\ref{one}d), in which the photon interacts
through its partonic content, and the direct process
(Fig.~\ref{one}e), in which the photon interacts as a point-like
particle. The cross section for the process $\eprr$ is given by the
convolution of the flux of photons in the electron, the parton
densities in the proton, the parton densities in the photon and the
subprocess cross section:

$$d\sigma_{ep\rightarrow {\rm jet\ jet}}=\sum_{i,j}\int_0^1 dy\ f_{\gamma /e}(y)\int_0^1 dx_{\gamma}\ f_{i/\gamma}(x_{\gamma},\mu_{F_{\gamma}})\int_0^1 dx_p\ f_{j/p}(x_p,\mu_{F_p})\ d\hat\sigma_{i(\gamma)j}(i(\gamma)j\rightarrow {\rm jet\ jet}),$$
where $f_{\gamma /e}$ is the flux of photons in the electron,
usually estimated using the Weizs\" acker-Williams approximation, $y$ is
the inelasticity variable, $f_{j/p}$ are the parton densities in the
proton, determined from global fits, $x_p$ is the proton momentum
taken by the interacting parton, $\mu_{F_p}$ is the proton
factorisation scale, $f_{i/\gamma}$ are parton densities in the photon,
determined from fits to deep inelastic $e\gamma$ data, $x_{\gamma}$ is
the photon momentum taken by the interacting parton,
$\mu_{F_{\gamma}}$ is the photon factorisation scale, and
$d\hat\sigma_{i(\gamma)j}(i(\gamma)j\rightarrow {\rm jet\ jet})$ is
the subprocess cross section, which is calculable in pQCD at any
order.

All the data accumulated from HERA~\cite{zeusf2,h1f2} and
fixed-target~\cite{fixed} experiments have allowed a good
determination of the proton PDFs over a large phase space and the
evolution of the PDFs with the scale $\mu_{F_p}$ for large scales has
been successfully described by the DGLAP evolution
equations~\cite{dglap}. Measurements of jet production in NC
DIS~\cite{zeusjetnc,h1jetnc} and
photoproduction~\cite{zeusjetgp,h1jetgp} have provided accurate tests
of pQCD and a determination of the fundamental parameter of the
theory, $\as$. Most of these measurements refer to the production of
jets irrespective of their partonic origin --quarks or gluons-- and,
therefore, have provided general tests of the partonic structure of
the short-distance process and of combinations of the proton and/or
photon PDFs. The identification of quark- and gluon-initiated jets
would allow more stringent tests of the QCD predictions.

At high scales, calculations using the DGLAP evolution equations have
been found to give a good description of the data up to
next-to-leading order (NLO). Therefore, by fitting the data with these
calculations, it has been possible to extract accurate values of $\as$
and the gluon density of the proton. However, for scales of 
$\etjet\sim Q$, where $\etjet$ is the jet transverse energy, large
values of the jet pseudorapidity, $\etajet$, and low values of $x$
discrepancies between the data and the NLO calculations have been
observed. This could indicate a breakdown of the DGLAP evolution and
the onset of BFKL~\cite{bfkl} effects. These discrepancies could also
be explained by assigning a partonic structure to the exchanged
virtual photon or a large contribution of higher-order effects at low
$\q2$.

This report includes recent results of the studies on parton
evolution at low $x$ from the H1 Collaboration, namely forward-jet
cross sections as a function of $x$ and the azimuthal correlation
between the hard jets in dijet events in NC DIS, and the tests of pQCD
at high scales from the ZEUS Collaboration, namely dijet and three-jet
cross sections in NC DIS, measurements of the internal structure of jets
in photoproduction and NC DIS and the study of the substructure
dependence of jet cross sections in photoproduction.

\section{Parton evolution at low Bjorken-$x$}
\begin{figure}[ht]
\setlength{\unitlength}{1.0cm}
\begin{picture} (10.0,7.0)
\put (5.0,0.0){\epsfig{figure=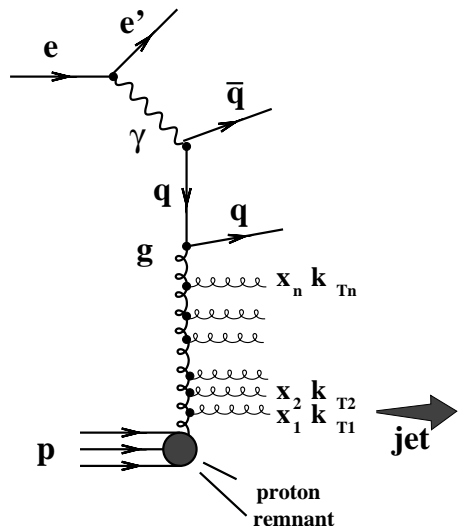,width=6.0cm}}
\end{picture}
\caption{Example of Feynman diagram for forward jet production in NC
  DIS at low $x$.
  \label{two}}
\end{figure}

To leading logarithm accuracy, the DGLAP evolution is equivalent to
the exchange of a parton cascade with the exchanged partons strongly
ordered in virtuality (Fig.~\ref{two}). The DGLAP equations sum the
leading powers of $\as\log{\q2}$ in the region of strongly-ordered
transverse momenta. However, DGLAP evolution is expected to breakdown
at low $x$ since only leading logarithms in $\q2$ are resummed and
contributions from $\log 1/x$ are neglected. These terms need to be
taken into account since when $\log\q2\ll\log 1/x$, terms proportional
to $\as\log 1/x$ become important.

Several theoretical approaches exist which account for low-$x$ effects
not incorporated into the DGLAP evolution. They are: (i) the BFKL
evolution which resums large $\log (1/x)$ terms to all orders, this
approach works at very low $x$ and presents no $\kt$ ordering, the
integration is taken over the full $\kt$ phase scape of the gluons and
the calculations make use of off-shell matrix elements together with
unintegrated PDFs; (ii) the CCFM~\cite{ccfm} evolution provides
angular-ordered parton emission and works for low and larger $x$, it
is equivalent to BFKL for $x\rightarrow 0$ and to DGLAP at large $x$;
(iii) the virtual-photon structure mimicks higher-order QCD effects at
low $x$ by introducing a second $\kt$-ordered parton cascade on the
photon side~\cite{resolved} \`{a} la DGLAP, the resolved contribution
is expected to contribute for $(\etjet)^2>\q2$ and suppressed with
increasing $\q2$.

There exist several programs to calculate pQCD predictions of jet
cross sections in NC DIS, eg {\sc Disent}~\cite{disent} and 
{\sc Nlojet}~\cite{nlojet}. These programs use the DGLAP evolution
equations. Higher-order effects can be mimicked by the parton-shower
approach in leading-logarithm Monte Carlo models like 
{\sc Rapgap}~\cite{rapgap}, which includes direct or direct plus 
resolved processes in virtual-photon interactions. Low-$x$ effects not
included in the DGLAP evolution are incorporated in Monte Carlo models
like {\sc Cascade}~\cite{cascade} and {\sc Ariadne}~\cite{ariadne}. 
{\sc Cascade} is based on the $\kt$-factorised unintegrated parton
distributions and follows the CCFM evolution, whereas {\sc Ariadne}
generates non-$\kt$ ordered parton cascades based on the color dipole
model (CDM).

Experimentally, deviations from the DGLAP evolution can be expected at
low $x$ and forward-jet rapidity since parton emission along the
exchanged gluon ladder (see Fig.~\ref{two}) increases with decreasing
$x$. Another method to obtain evidence for DGLAP breakdown is to study
the azimuthal correlation between the two hardest jets. In DGLAP,
partons entering the hard process with negligible $\kt$ produce a
back-to-back configuration at LO. Values of
$\Delta\phi\!<\!180^{\circ}$ occur in DGLAP due to higher-order QCD
effects. In models which predict a significant proportion of partons
entering the hard process with large $\kt$, the number of events with
small $\Delta\phi$ will increase.

\subsection{$x$ dependence of the forward-jet cross section}

The forward-jet cross section has been measured~\cite{h1prel} for jets
identified with the $\kt$ cluster algorithm in the longitudinally
inclusive mode in the laboratory frame. Events with at least one jet of
transverse energy $\etlab>3.5$ GeV and $1.7<\etalab<2.8$ were
selected. The events are required to fulfill the additional
conditions: (i) $x_{\rm jet}=E_{\rm jet}/E_p>0.035$,
where $E_p$ is the proton-beam energy, and (ii) $0.5<(\etjet)^2/\q2<5$,
following the proposal of Mueller and Navelet~\cite{navelet} such as
to allow for a large-level arm of evolution in $x$ and to restrict the
evolution in $\q2$. The measurements were made in the kinematic region
given by $5<\q2<85$ \g2\ and $0.0001<x<0.004$.

\begin{figure}[ht]
\setlength{\unitlength}{1.0cm}
\begin{picture} (10.0,9.0)
\put (0.0,0.0){\epsfig{figure=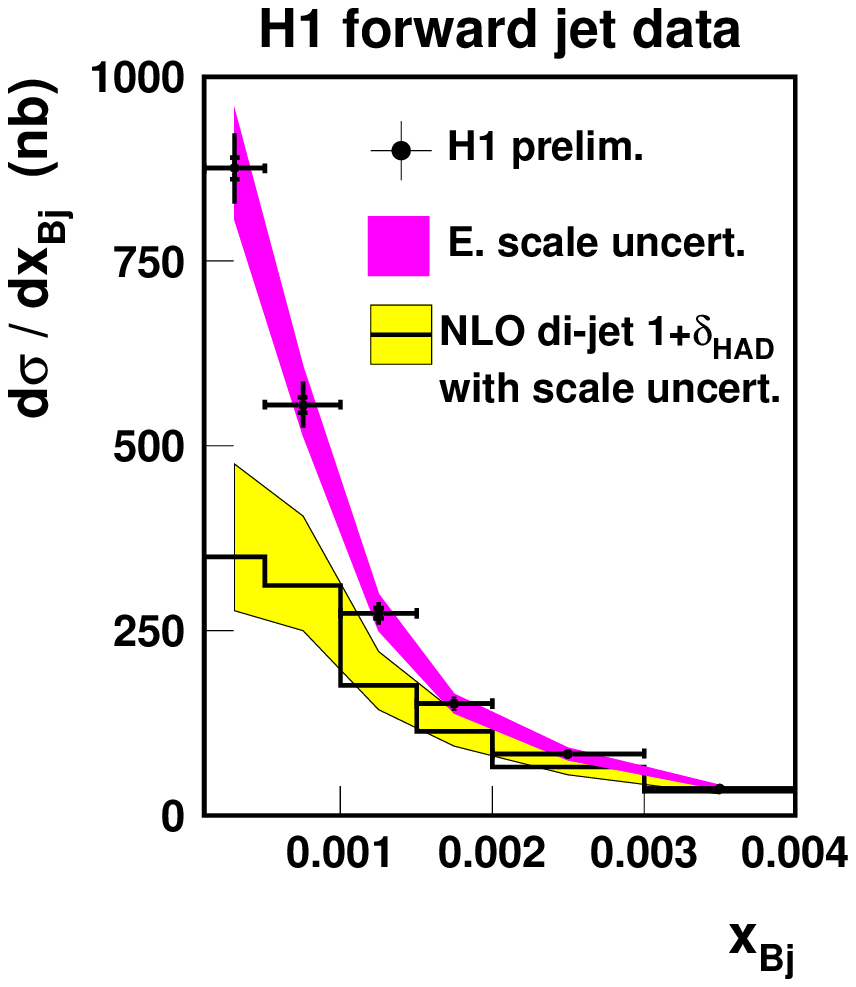,width=7.5cm}}
\put (8.0,0.0){\epsfig{figure=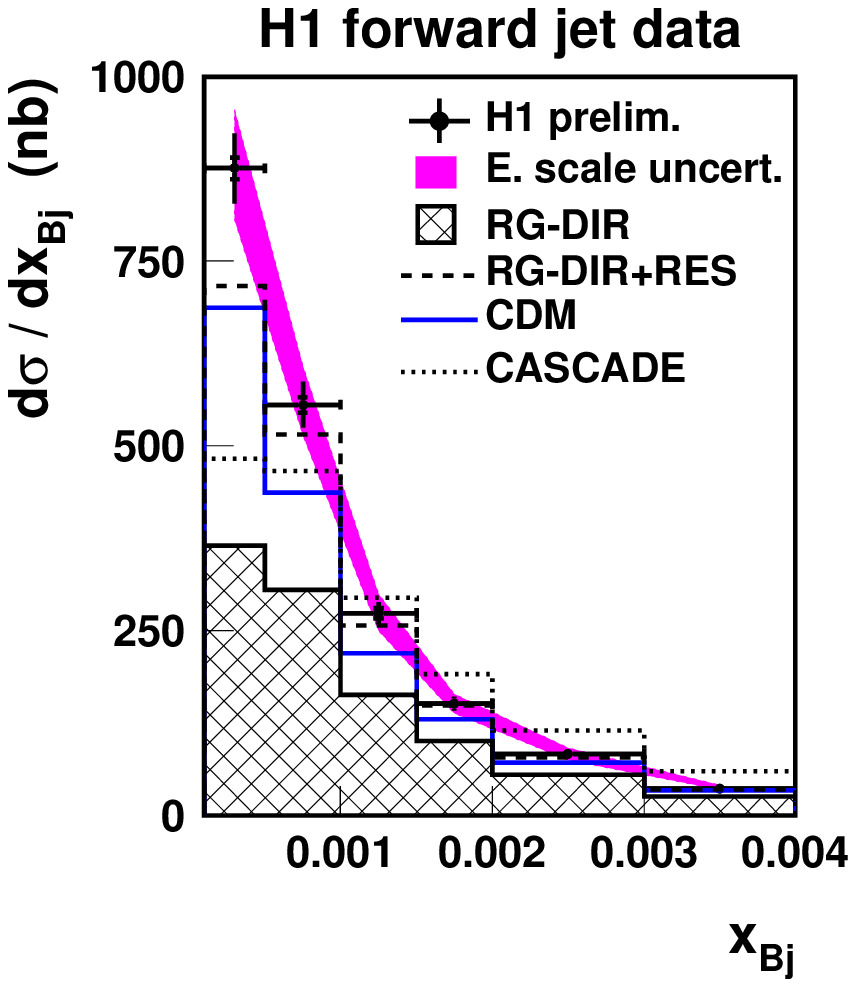,width=7.5cm}}
\put (3.5,0.0){\bf\small (a)}
\put (11.7,0.0){\bf\small (b)}
\end{picture}
\caption{Measured forward-jet cross section
  (dots)~\protect\cite{h1prel} as a function of Bjorken $x$. For
  comparison, the predictions of (a) {\sc Disent} and (b) {\sc Rapgap}
  (RG), {\sc Ariadne} (CDM) and {\sc Cascade} are also included.
  \label{three}}
\end{figure}

Figure~\ref{three} shows the forward-jet cross section as a
function of $x$. The measured cross section rises with decreasing $x$.
The NLO calculation corrected for hadronisation effects obtained using
the program {\sc Disent} with the renormalisation scale $\mu^2_R=45$
\g2, the average $(\etjet)^2$ of the data, and the CTEQ6
parametrisations of the proton PDFs, is compared to the
data in Fig.~\ref{three}a. The measured cross section is well
described by the prediction for large values of $x$, but at low $x$
values there is a large excess of data with respect to the
calculation. This prediction is based on DGLAP evolution and so it is 
not expected to work in this region of phase space.

Figure~\ref{three}b shows the comparison with the predictions of
different leading-logarithm parton-shower Monte Carlo models. The
prediction of {\sc Rapgap} including only direct processes is similar
to the NLO calculation. Once the contribution from resolved processes
is included, a better description of the data is obtained for $x$ down
to $0.001$. The {\sc Cascade} prediction, based on the CCFM evolution,
does not reproduce the shape of the data, whereas the prediction from
CDM describes the data for $x>0.0015$. In conclusion, no model can
describe the sharp rise of the data at very low $x$, which remains a
challenge and demands improved approximations to QCD in that region of
phase space.

\subsection{Azimuthal jet separation}

\begin{figure}[ht]
\setlength{\unitlength}{1.0cm}
\begin{picture} (10.0,10.0)
\put (0.0,0.2){\epsfig{figure=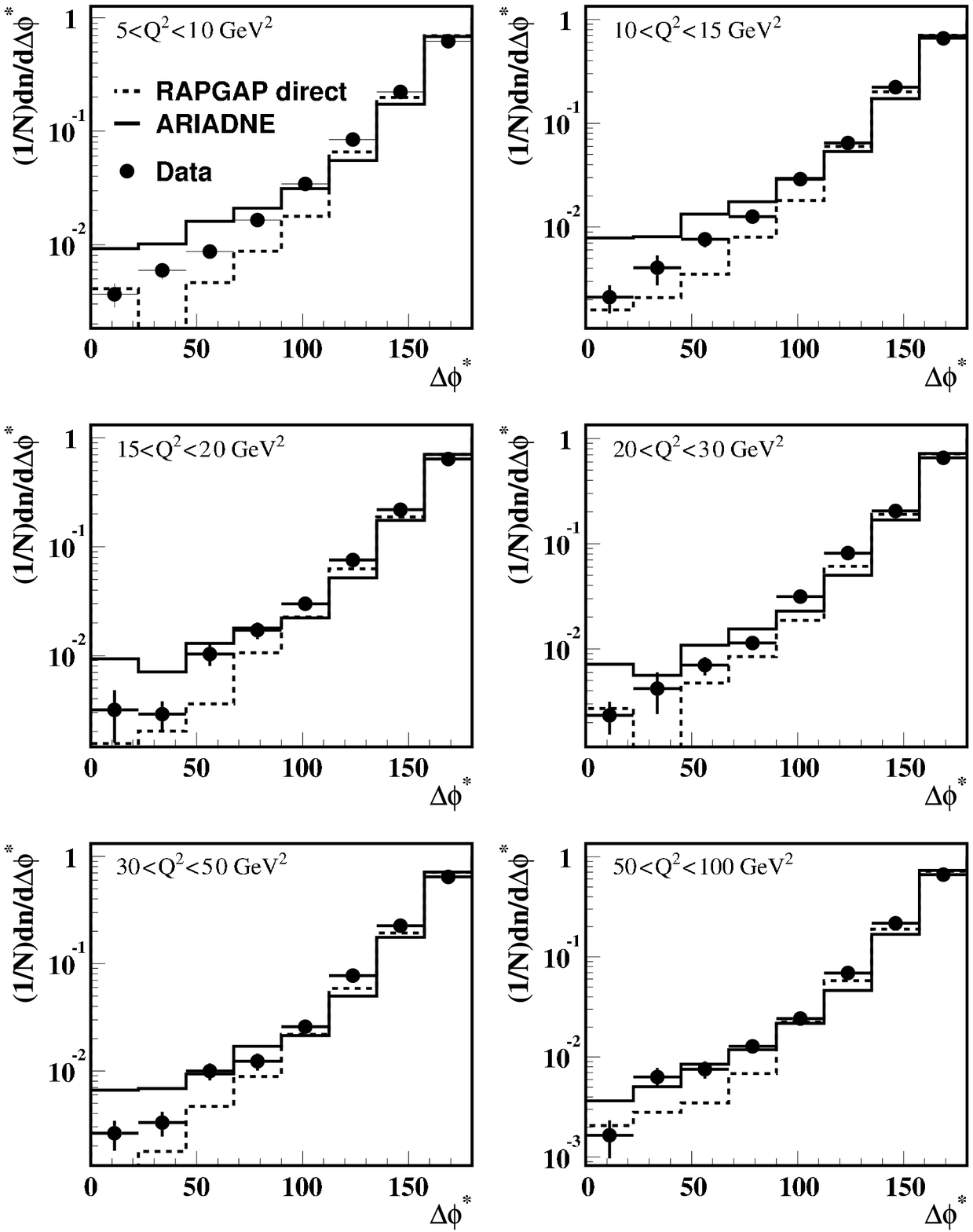,width=7.5cm}}
\put (9.0,0.0){\epsfig{figure=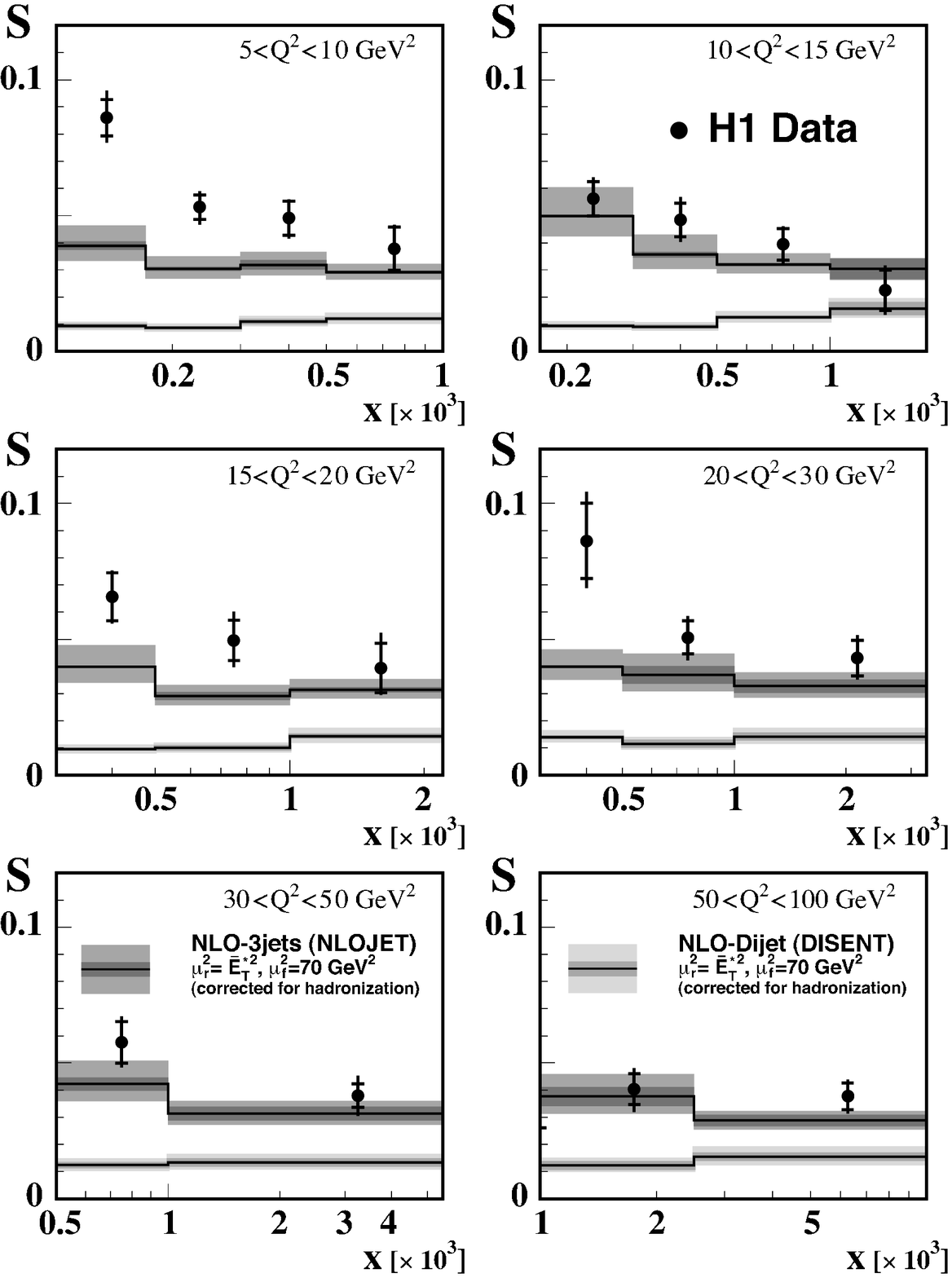,width=7.5cm}}
\put (3.5,-0.2){\bf\small (a)}
\put (12.7,-0.2){\bf\small (b)}
\end{picture}
\caption{(a) Measured $\Delta\phi^*$ distribution in different
  regions of $\q2$ (dots)~\protect\cite{dijh1}. For comparison, the
  predictions of {\sc Rapgap} and {\sc Ariadne} are also included. (b)
  Measured fraction $S$ as a function of $x$ in different $\q2$
  regions (dots)~\protect\cite{dijh1}. For comparison, the predictions
  of the fixed-order QCD calculations from {\sc Disent} and 
  {\sc Nlojet} are also included.
  \label{four}}
\end{figure}

The azimuthal correlation between the two hard jets in dijet events
has been measured~\cite{dijh1} using the $\kt$-cluster algorithm in
the longitudinally inclusive mode in the $\gamma^* p$ centre-of-mass
frame. The measurements were made in the kinematic region given by
$5<\q2<100$ \g2\ and $10^{-4}<x<10^{-2}$. The cross sections refer to
jets of $E_T^*>5$ GeV, $-1<\etalab<2.5$ and $E_{T,{\rm max}}^*>7$ GeV,
where $E_T^*$ is the jet transverse energy in the $\gamma^*p$
centre-of-mass frame. Figure~\ref{four}a shows the measured
distribution as a function of the azimuthal separation in different
$\q2$ regions. A significant fraction of events is observed at small
azimuthal separations. Since a measurement of a multi-differential
cross section as a function of $x$, $\q2$ and $\Delta\phi^*$ would be
very difficult due to large migrations, the fraction of the number of
dijet events with an azimuthal separation between $0$ and $\alpha$,
where $\alpha$ was taken as $\alpha=\frac{2}{3}\pi$, was measured
instead. The fraction $S$, defined as
$$S=\frac{\int_0^{\alpha} N_{\rm 2jet}(\Delta\phi^*,x,\q2)d\Delta\phi^*}{\int_0^{\pi} N_{\rm 2jet}(\Delta\phi^*,x,\q2)d\Delta\phi^*},$$
is better suited to test small-$x$ effects than a triple-differential
cross section.

\begin{figure}[ht]
\setlength{\unitlength}{1.0cm}
\begin{picture} (10.0,10.0)
\put (0.0,0.1){\epsfig{figure=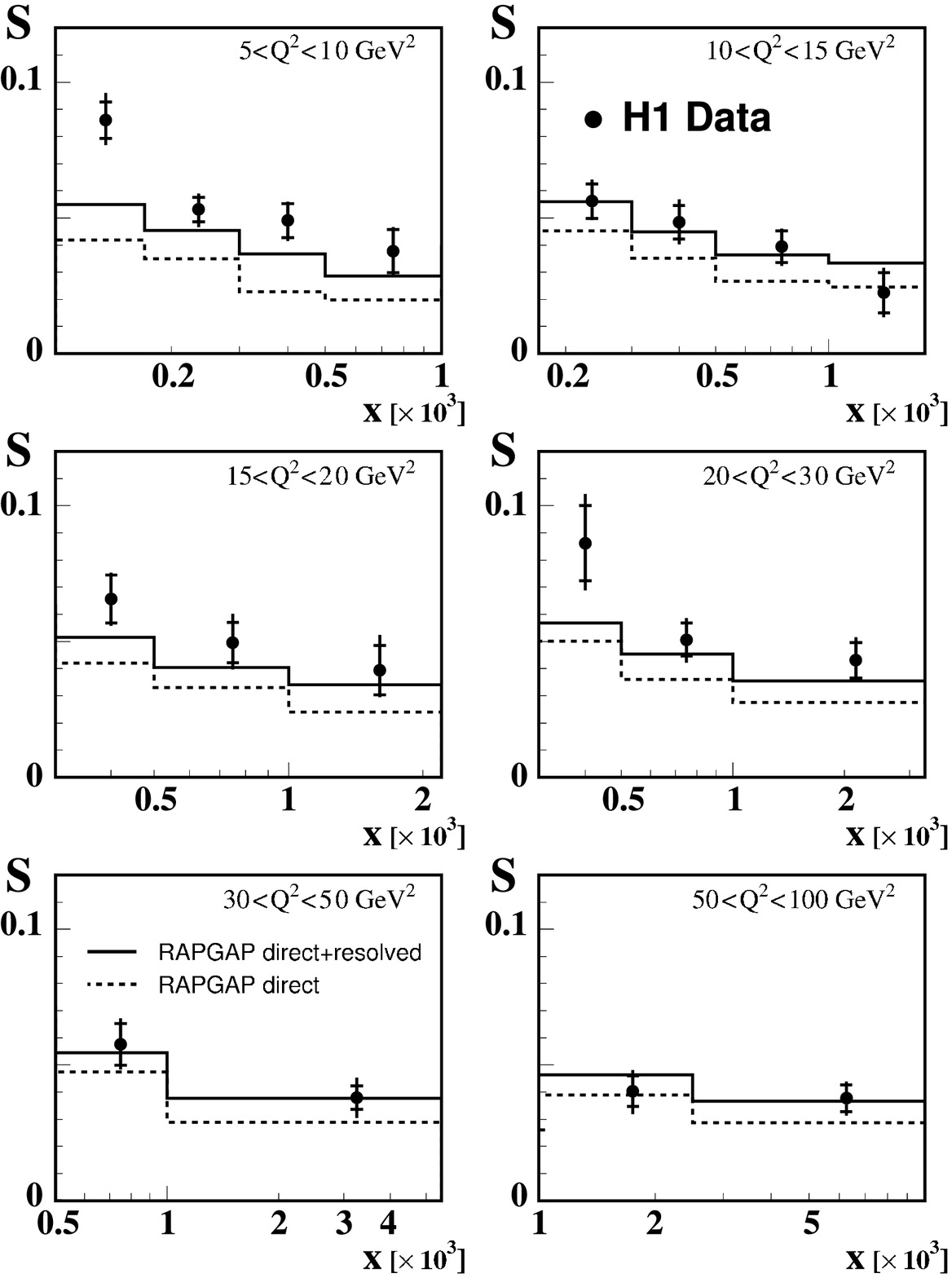,width=7.5cm}}
\put (9.0,0.0){\epsfig{figure=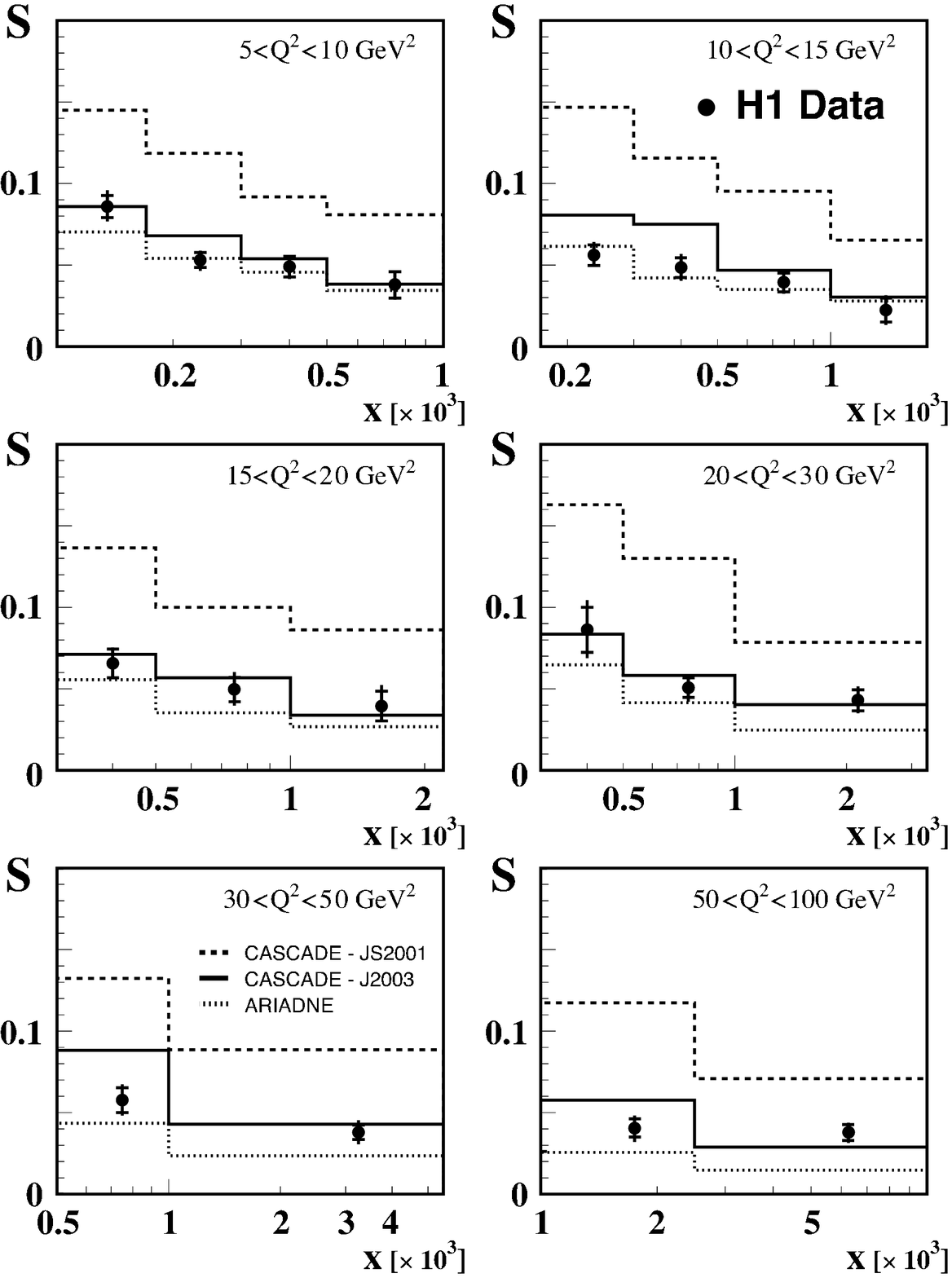,width=7.5cm}}
\put (3.5,-0.2){\bf\small (a)}
\put (12.7,-0.2){\bf\small (b)}
\end{picture}
\caption{Measured fraction $S$ as a function of $x$ in different $\q2$
  regions (dots)~\protect\cite{dijh1}. For comparison, the predictions
  from (a) {\sc Rapgap} and (b) {\sc Cascade} and {\sc Ariadne} are
  also included.
  \label{five}}
\end{figure}

The measured fraction $S$ as a function of $x$ in different
regions of $\q2$ is presented in Fig.~\ref{four}b. The data rise
towards low $x$ values, especially at low $\q2$. The predictions
from {\sc Disent}, which contain the lowest-order contribution
to $S$, are several standard deviations below the data and show
no dependence with $x$. On the other hand, the predictions of {\sc
  Nlojet}, which incorporate NLO corrections to $S$, provide a good
description of the data at large $\q2$ and large $x$. However, they
fail to describe the increase of the data towards low $x$ values,
especially at low $\q2$. This shows the need to incorporate NLO
corrections to the calculations.

Higher-order effects can be mimicked by the parton shower approach in
Monte Carlo models. Figure~\ref{five}a shows the comparison with the
predictions of {\sc Rapgap} with direct only and resolved plus direct
processes. A good description of the data is obtained at large $\q2$
and large $x$. However, there is a failure to describe the strong rise
of the data towards low $x$, especially at low $\q2$, even when
including a possible contribution from resolved virtual-photon
processes, though the description in other regions is improved.

If the observed discrepancies are due to the influence of non-ordered
parton emission, models based on the CDM or the CCFM evolution could
provide a better description of the data. Figure~\ref{five}b shows the
data compared with {\sc Ariadne} and two predictions of {\sc Cascade},
which use different sets of unintegrated parton distributions. These
sets differ in the way the small-$\kt$ region is treated: in Jung2003
the full splitting function, i.e. including the non-singular term, is
used in contrast to JS2001, for which only the singular term was
considered. The predictions of {\sc Ariadne} give a good description
of the data at low $x$ and $\q2$, but fail to describe the data at
high $\q2$. The predictions of {\sc Cascade} using
JS2001~\cite{cascade} lie significantly above the data in all $x$ and
$\q2$ regions, whereas those using Jung2003~\cite{jung2003} are closer
to the data. Therefore, the measurement of the fraction $S$ is
sensitive to and can be used to gain information on the unintegrated
parton distributions.

\section{Multi-jet production in NC DIS}
\begin{figure}[ht]
\setlength{\unitlength}{1.0cm}
\begin{picture} (10.0,6.3)
\put (1.0,0.2){\epsfig{figure=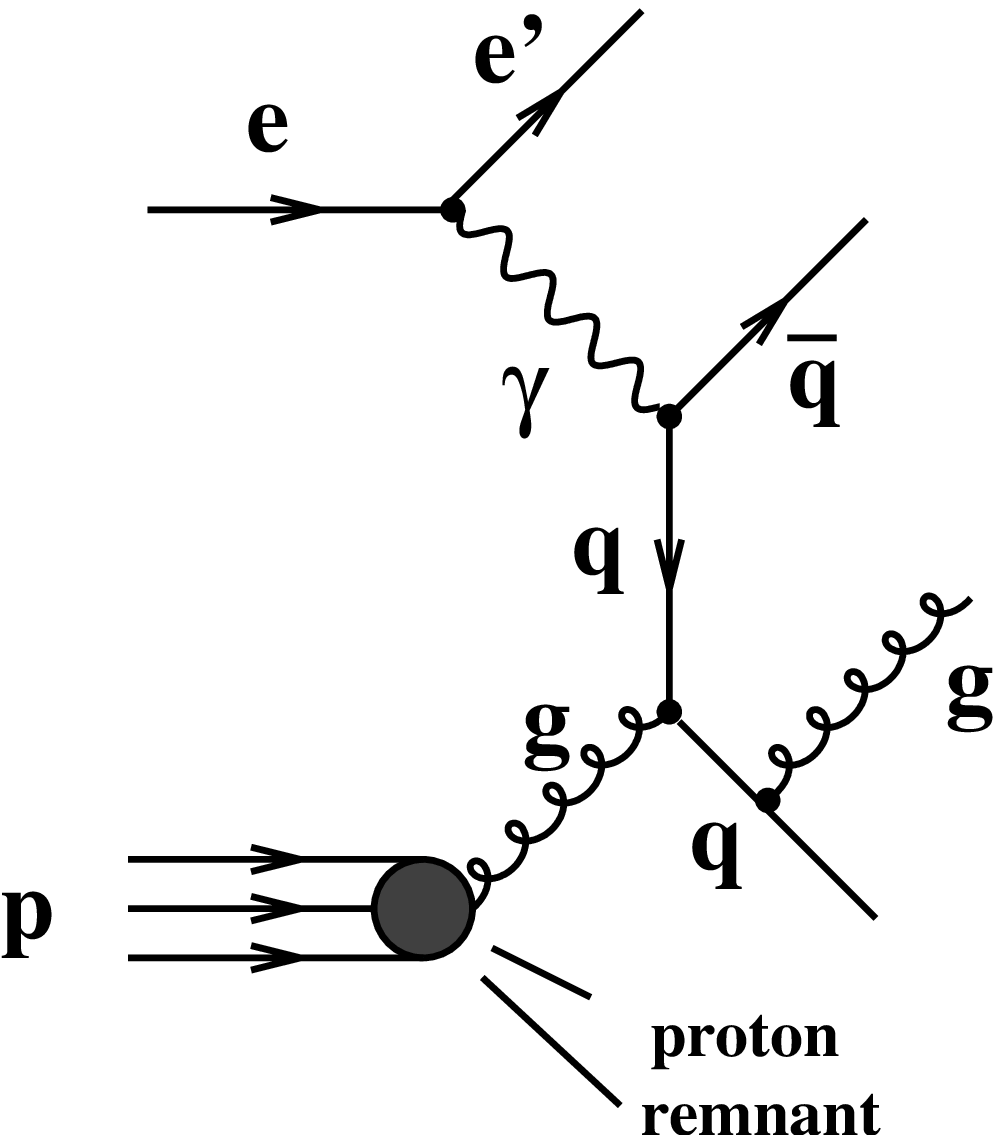,width=5.0cm}}
\put (9.5,0.2){\epsfig{figure=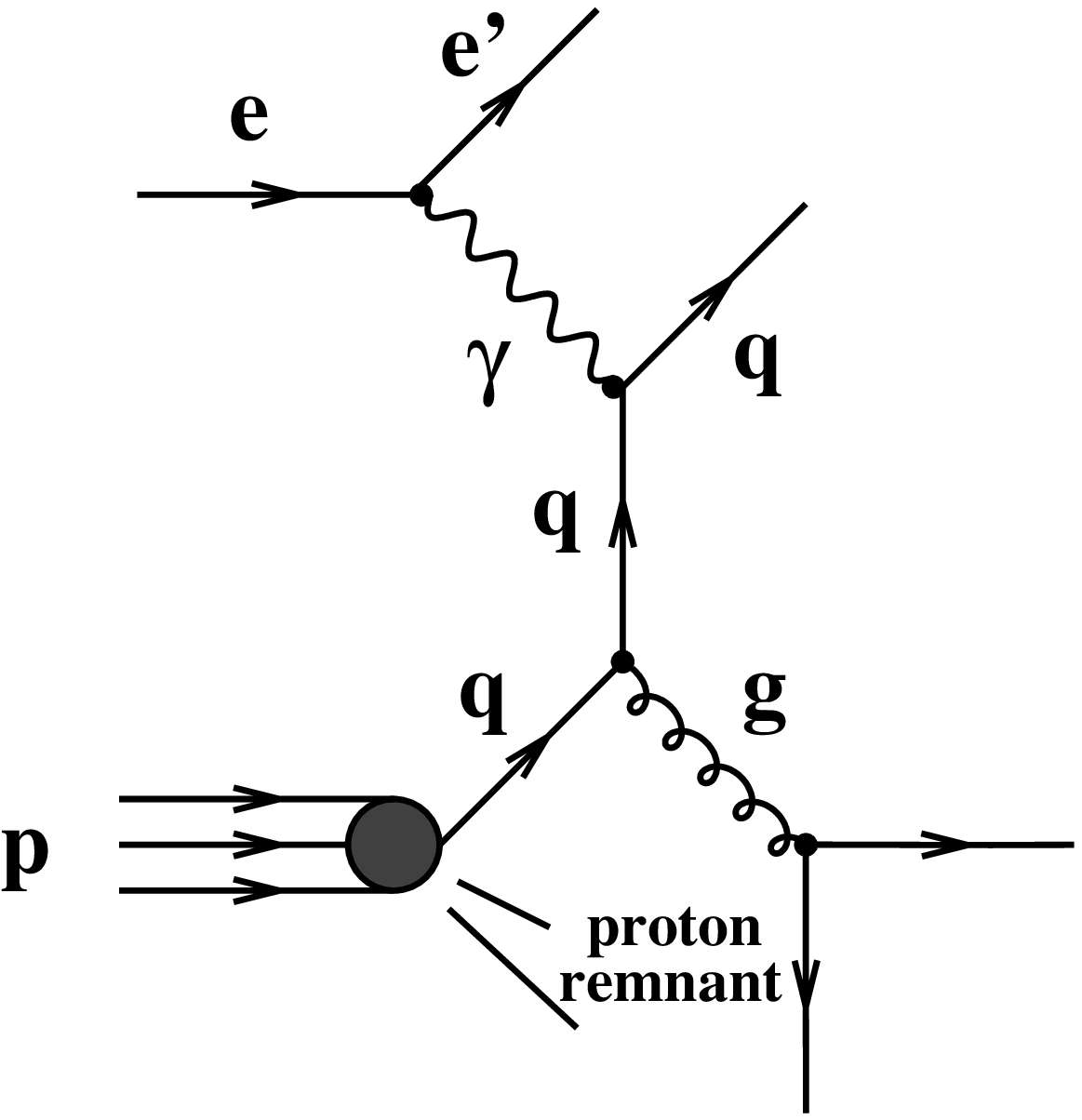,width=5.5cm}}
\put (3.5,-0.2){\bf\small (a)}
\put (12.5,-0.2){\bf\small (b)}
\end{picture}
\caption{Examples of Feynman diagrams for three-jet production in NC
  DIS: (a) gluon brehmstrahlung from a quark and (b)
  gluon splitting into a $\qq$ pair.
  \label{six}}
\end{figure}

Three-jet production in NC DIS provides a test of pQCD directly beyond
LO since the since the lowest-order contribution is proportional to 
$\as^2$. Three-jet events arise from additional gluon
brehmstrahlung or splitting of a gluon into a $\qq$ pair
(Fig.~\ref{six}).

\begin{figure}[ht]
\setlength{\unitlength}{1.0cm}
\begin{picture} (10.0,9.0)
\put (0.0,0.0){\epsfig{figure=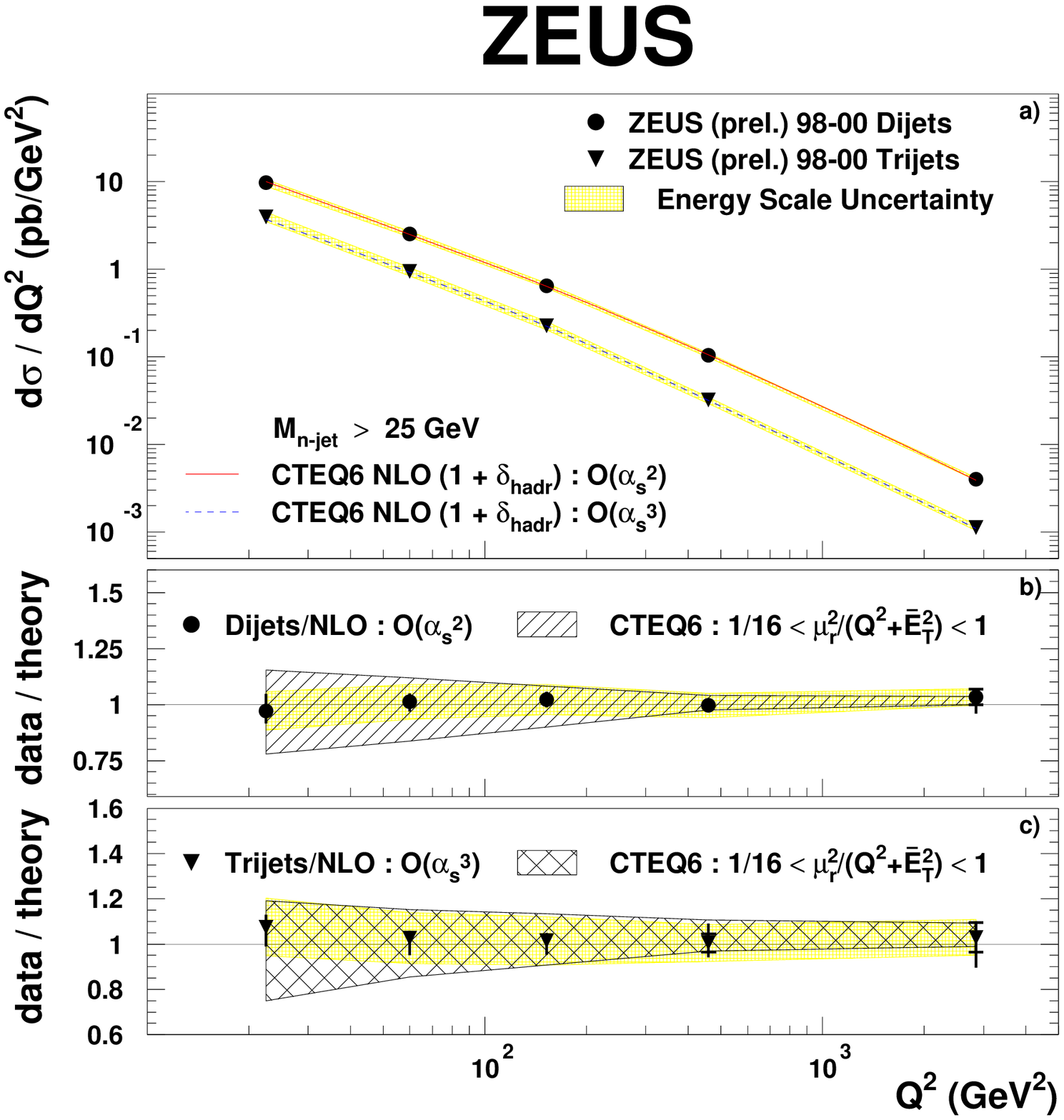,width=9.0cm}}
\put (8.5,0.0){\epsfig{figure=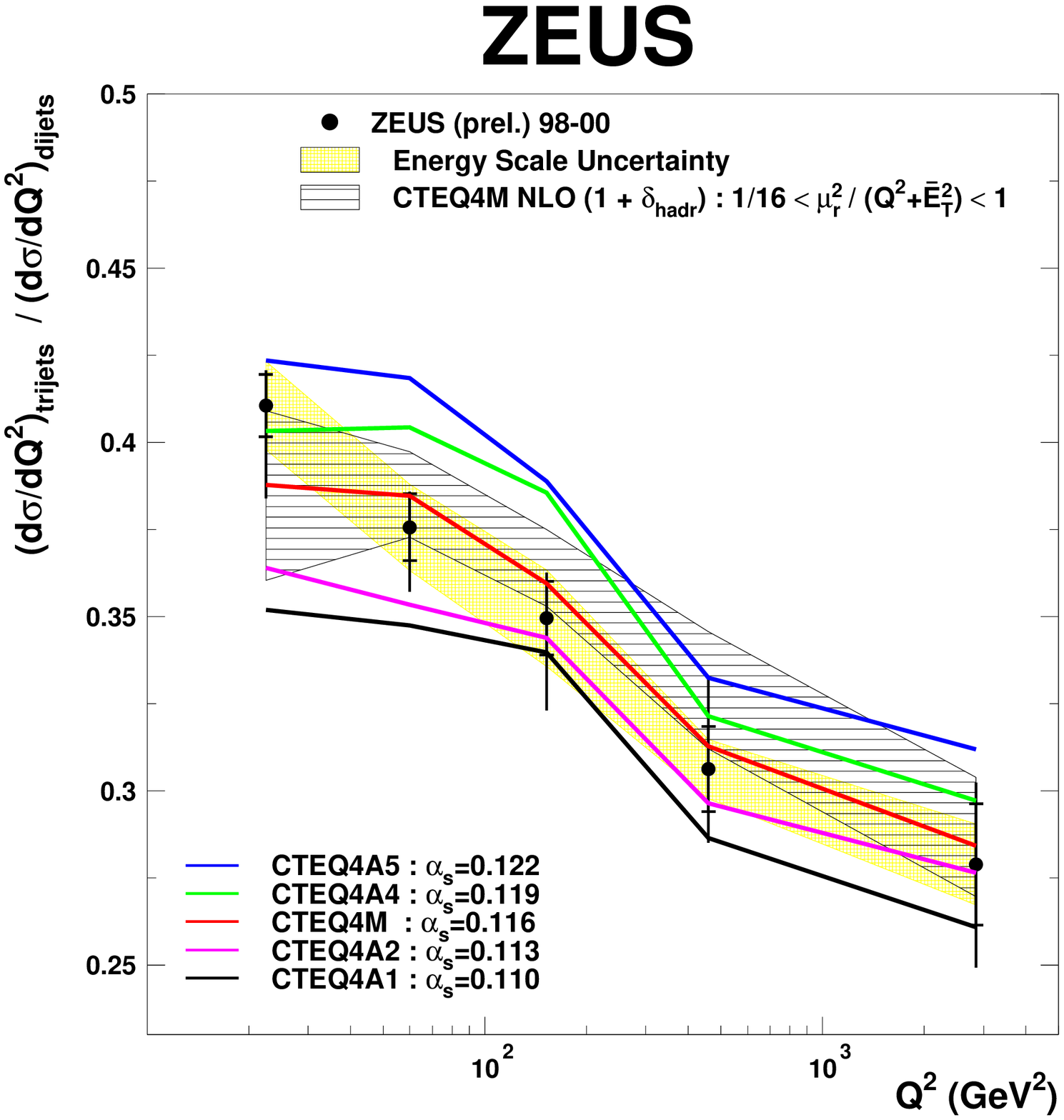,width=9.0cm}}
\put (3.5,-0.2){\bf\small (a)}
\put (12.5,-0.2){\bf\small (b)}
\end{picture}
\caption{(a) Measured dijet and three-jet cross sections in NC DIS as a
  function of $\q2$ (dots)~\protect\cite{multi}. (b) Measured ratio of
  the three-jet to the dijet cross section as a function of $\q2$
  (dots)~\protect\cite{multi}. For comparison, the predictions of {\sc
    Nlojet} are also included in both figures.
  \label{seven}}
\end{figure}

The dijet (three-jet) cross section has been measured~\cite{multi} 
using the $\kt$ cluster algorithm in the
longitudinally invariant mode in the Breit frame for events with at
least two (three) jets of $\etjb>5$ GeV and $-1<\etalab<2.5$, and with
dijet (three-jet) invariant masses in excess of $25$ GeV; the
kinematic region is defined by $10<\q2<5000$ \g2. Figure~\ref{seven}a
shows the dijet and three-jet cross sections as a function of
$\q2$. The data are compared to the predictions of {\sc Nlojet} up to
${\cal O}(\as^2)$ and ${\cal O}(\as^3)$, respectively, using
$\mu_R^2=\q2+(\bar\etjb)^2$, where $\bar\etjb$ is the average jet
transverse energy of the two (three) jets, and the factorisation scale
$\mu_F=Q$. The CTEQ6 parametrisations of the proton PDFs have been
used for the proton PDFs. The measured cross sections are well
described by the predictions.

The $\q2$ dependence of the ratio of the three-jet to the dijet cross
section has been studied (see Fig.~\ref{seven}b). Many experimental
and theoretical uncertainties cancel in the ratio and therefore this
observable provides a more accurate test of color dynamics, especially
at low $\q2$ since the theoretical uncertainty of the ratio is of the 
same order as at higher $\q2$. The calculations from {\sc Nlojet},
using the five sets of the CTEQ4 "A-series" of proton PDFs, give a
good description of the data. The predictions for different values of
$\as$ show the sensitivity of this observable to the value of $\asz$
assumed in the calculations. A value of
$$\asmz{0.1179}{0.0013}{0.0046}{0.0028}{0.0047}{0.0061}$$
has been extracted from the ratio, which is in good agreement with the
world average, $0.1182\pm 0.0027$~\cite{bethke} and other
determinations of $\as$ from HERA data.

\section{Jet substructure and the dynamics of quarks and gluons}
The internal structure of a jet depends mainly on the type of primary
parton --quark or gluon-- from which it originated and to a lesser
extent on the particular hard scattering process. QCD predicts that at
sufficiently high $\etjet$, where fragmentation effects become
negligible, the jet structure is driven by gluon emission off the
primary parton and is then calculable in pQCD. QCD also predicts that
gluon jets are broader than quark jets due to the larger colour charge
of the gluon. This prediction has been confirmed at LEP in
measurements of the internal structure of jets~\cite{lep}.

The internal structure of jets has been studied by means of the jet
shape. The integrated jet shape is defined as the fraction of the jet
transverse energy that lies inside a cone in the $\etaphi$ plane of
radius $r$ concentric with the jet axis, using only those particles
belonging to the jet~\cite{jetshape}:

$$\psi(r)=\frac{E_T(r)}{\etjet},$$
where $E_T(r)$ is the transverse energy within the given cone of
radius $r$. The mean integrated jet shape, $\langle\psi(r)\rangle$, is
defined as the averaged fraction of the jet transverse energy inside
the cone of radius $r$:

$$\langle\psi(r)\rangle=\frac{1}{N_{\rm jets}}\sum_{\rm jets}\frac{E_T(r)}{\etjet},$$
where $N_{\rm jets}$ is the total number of jets in the sample.

The jets have been identified using
the $\kt$ cluster algorithm in the longitudinally inclusive mode and
selected according to $\etjet>17$ GeV and $\etar$. The kinematic
region in the photoproduction sample is defined by $\q2<1$ \g2\ and
\wrn, where $\wgp$ is the $\gp$ centre-of-mass energy, and in the NC
DIS sample by $\q2>125$ \g2.

\begin{figure}[ht]
\setlength{\unitlength}{1.0cm}
\begin{picture} (10.0,8.5)
\put (-0.5,0.0){\epsfig{figure=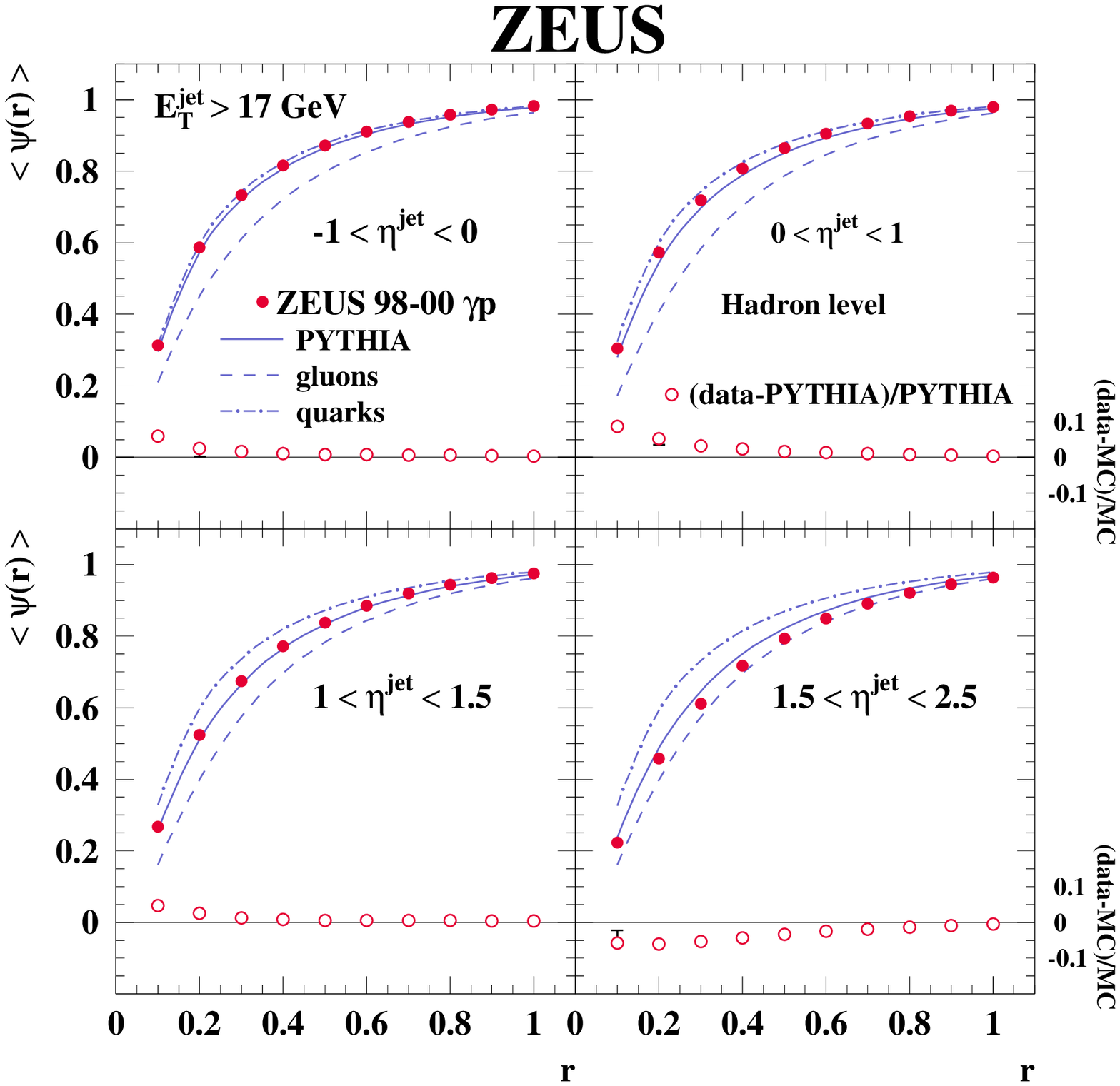,width=9.0cm}}
\put (8.5,0.0){\epsfig{figure=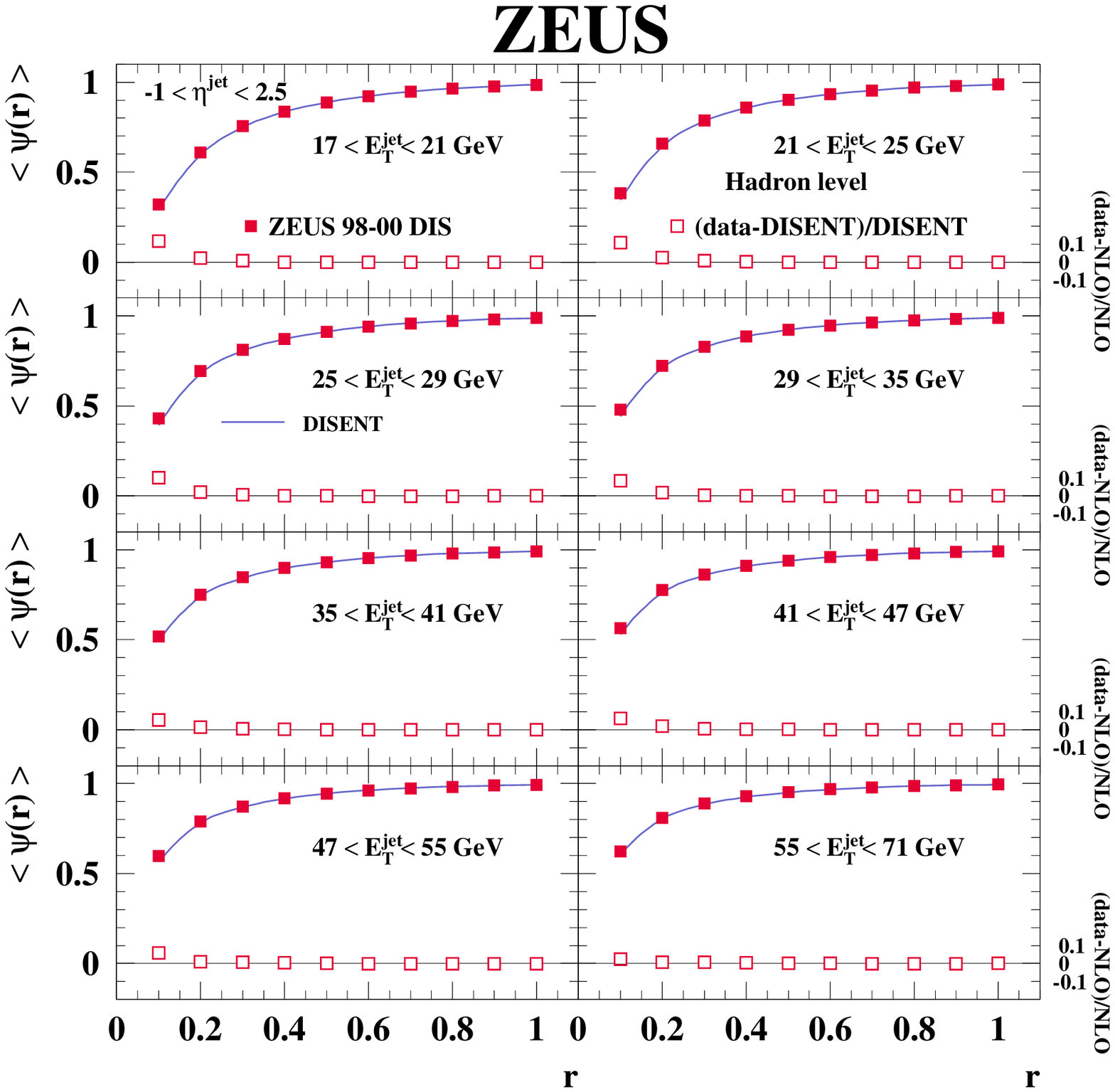,width=9.0cm}}
\put (3.6,-0.2){\bf\small (a)}
\put (12.6,-0.2){\bf\small (b)}
\end{picture}
\caption{(a) Measured mean integrated jet shape as a function of $r$ in
  different regions of $\etajet$ in photoproduction
  (dots)~\protect\cite{inczeus}. (b) Measured mean integrated jet
  shape as a function of $r$ in different $\etjet$ regions in NC DIS
  (squares)~\protect\cite{inczeus}. For comparison, the predictions of
  (a) {\sc Pythia} and (b) {\sc Disent} are also included.
  \label{eight}}
\end{figure}

\subsection{Jet-shape measurements}
The measured mean integrated jet shape as a function of $r$,
$\langle\psi(r)\rangle$, for different regions in $\etajet$ is shown
in Fig.~\ref{eight}a for the photoproduction
regime~\cite{inczeus}. The jets broaden as $\etajet$
increases. Leading-logarithm parton-shower predictions from {\sc
  Pythia}~\cite{pythia} for resolved plus direct processes and for
gluon- and quark-initiated jets are compared to the data in
Fig.~\ref{eight}a. The measured $\langle\psi(r)\rangle$ is reasonably
well described by the MC calculations of {\sc Pythia} for resolved and
direct processes for $-1<\etajet<1.5$, whereas for $1.5<\etajet<2.5$,
the measured jets are slightly broader than the predictions. From the
comparison with the predictions for gluon- and quark-initiated jets,
it is seen that the measured jets are quark-like for $-1<\etajet<0$
and become increasingly more gluon-like as $\etajet$ increases.

The measured $\langle\psi(r)\rangle$ for different regions of
$\etjet$ is shown in Fig.~\ref{eight}b for NC DIS events. The NLO QCD
calculations of $\langle\psi(r)\rangle$, corrected for hadronisation
and $\z0$-exchange effects, are compared to the data in the
figure. The NLO QCD calculations give a good description of
$\langle\psi(r)\rangle$ for $r\geq 0.2$; the fractional differences
between the measurements and the predictions are less than $0.2\%$ for
$r=0.5$.

\begin{figure}[ht]
\setlength{\unitlength}{1.0cm}
\begin{picture} (10.0,9.6)
\put (0.0,0.0){\epsfig{figure=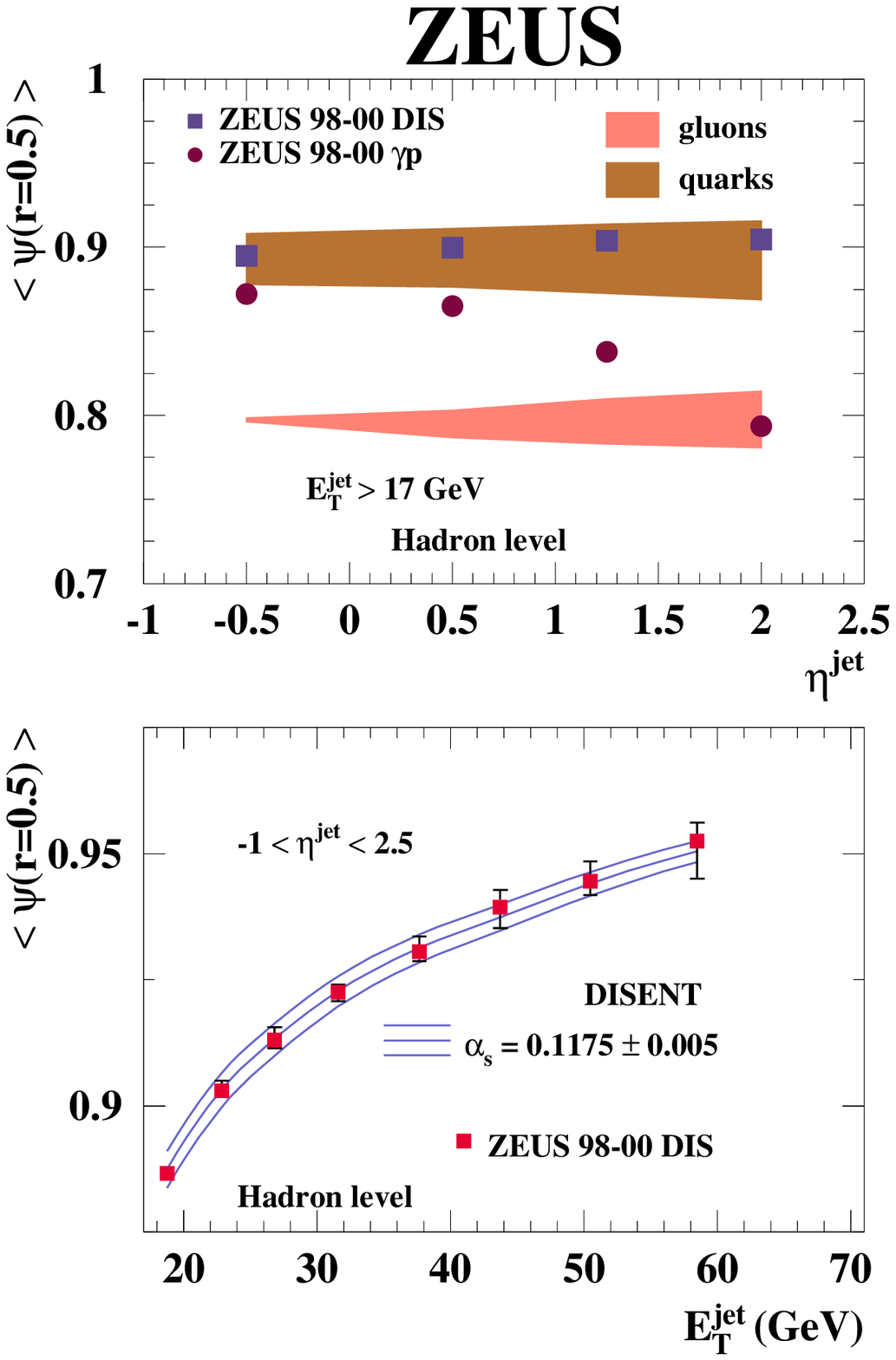,width=10.0cm}}
\put (8.5,0.0){\epsfig{figure=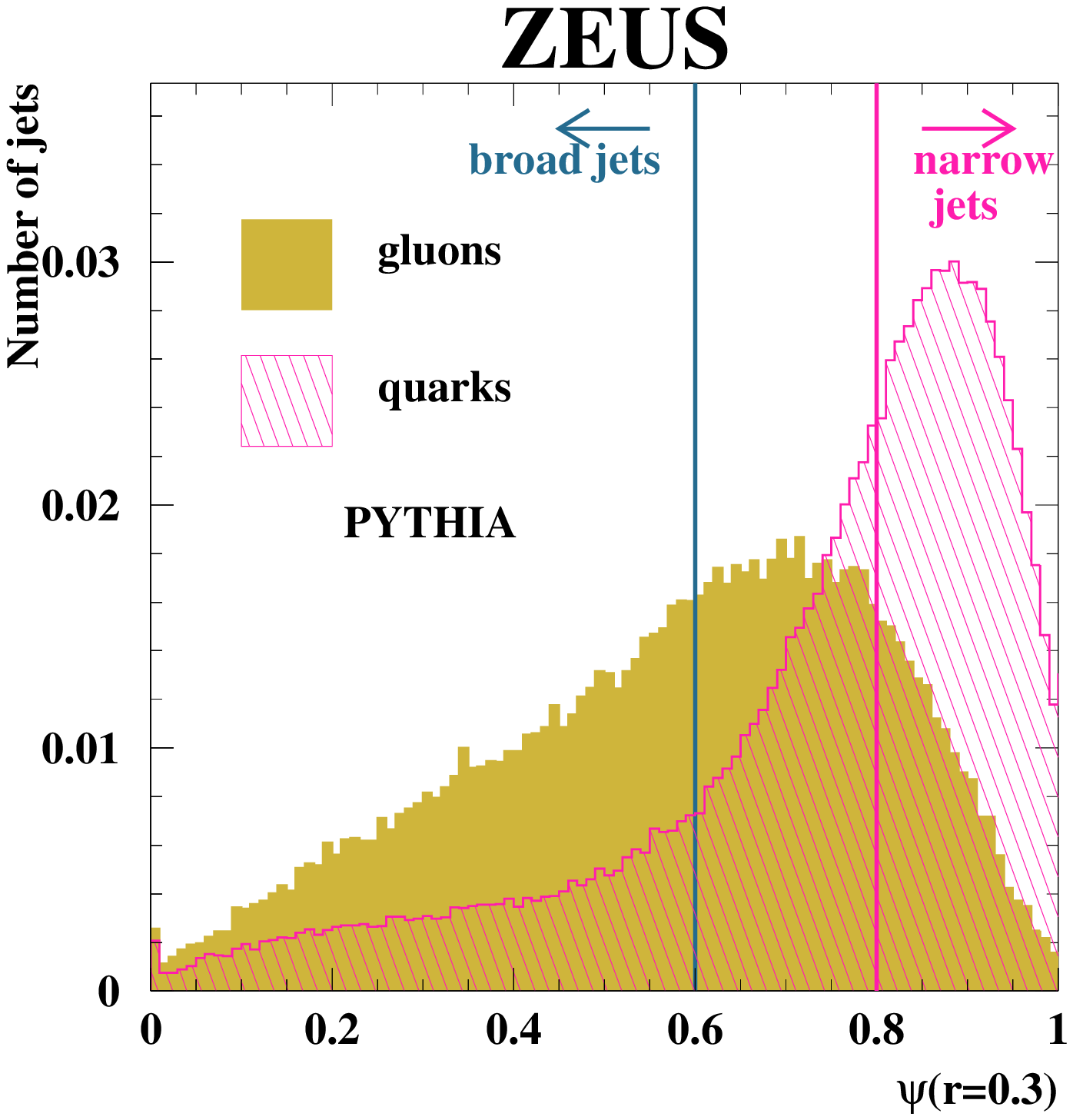,width=10.0cm}}
\put (0.5,8.5){\bf\small (a)}
\put (0.5,4.2){\bf\small (b)}
\put (12.5,1.0){\bf\small (c)}
\end{picture}
\caption{(a) Measured mean integrated jet shape at a fixed value of
  $r=0.5$ as a function of $\etajet$ in photoproduction (dots) and NC
  DIS (squares)~\protect\cite{inczeus}. For comparison, the Monte
  Carlo predictions for quark- and gluon-initiated jets are also
  included. (b) Measured mean integrated jet shape at a fixed value of
  $r=0.5$ as a function of $\etjet$ in NC
  DIS~\protect\cite{inczeus}. For comparison,the predictions of {\sc
    Disent} assuming three different values of $\as$ are also
  included. (c) Predicted integrated jet shape distributions from {\sc
    Pythia} at $r=0.3$ for quark- (hatched histogram) and
  gluon-initiated (shaded histogram) jets.
  \label{nine}}
\end{figure}

The quark and gluon content of the final state has been investigated
in more detail by studying the $\etajet$ dependence of the mean
integrated jet shape in photoproduction and NC DIS at a fixed value of
$r=0.5$ (see Fig.~\ref{nine}a). The jet shape at a fixed value of
$r=0.5$ decreases with increasing $\etajet$ for photoproduction,
whereas the jets in NC DIS show no dependence with $\etajet$. The
comparison of the data with the predictions for quark and gluon jets
shows that the NC DIS jets are consistent with being dominated by quark
jets, whereas the broadening of the jets in photoproduction is
consistent with an increase of the fraction of gluon jets as $\etajet$
increases. The dependence of the mean integrated jet shape for a fixed
value of $r=0.5$ as a function $\etjet$ (see Fig.\ref{nine}b) in NC
DIS shows that the jets become narrower as $\etjet$ increases. The
same is observed in photoproduction (not shown). The comparison of 
the data with the NLO calculations assuming different values of $\asz$
shows the sensitivity of this observable to the value of $\asz$. A
value of $\as$ of 
$$\asmz{0.1176}{0.0009}{0.0026}{0.0009}{0.0072}{0.0091}$$
was determined from this observable. This determination of
$\as$ has experimental uncertainties as small as those based on previous
measurements. However, the theoretical uncertainty is large and
dominated by terms beyond NLO. Further theoretical work on
higher-order contributions would allow an improved determination of
$\as$ from the integrated jet shape in DIS.

\subsection{Substructure dependence of jet cross sections}
The predictions of the Monte Carlo for the jet shape reproduce well
the data and show the expected differences for quark- and
gluon-initiated jets. These differences are used to select samples
enriched in quark and gluon jets to study in more detail the dynamics
of the hard subprocesses. The predicted shapes of the distribution in
$\psi(r=0.3)$ for quark and gluon jets are different, as shown in
Fig.~\ref{nine}c. A sample enriched in quark (or ``narrow'') jets was
selected by requiring an integrated jet shape above $0.8$ and a sample
enriched in gluon (or ``broad'') jets was selected by requiring an
integrated jet shape below $0.6$. {\sc Pythia} predicts a purity of
$57\%$ for gluons and $84\%$ for quarks, and the efficiencies are
$58\%$ for gluons and $51\%$ for quarks.

\begin{figure}[ht]
\setlength{\unitlength}{1.0cm}
\begin{picture} (10.0,8.0)
\put (-0.5,-1.0){\epsfig{figure=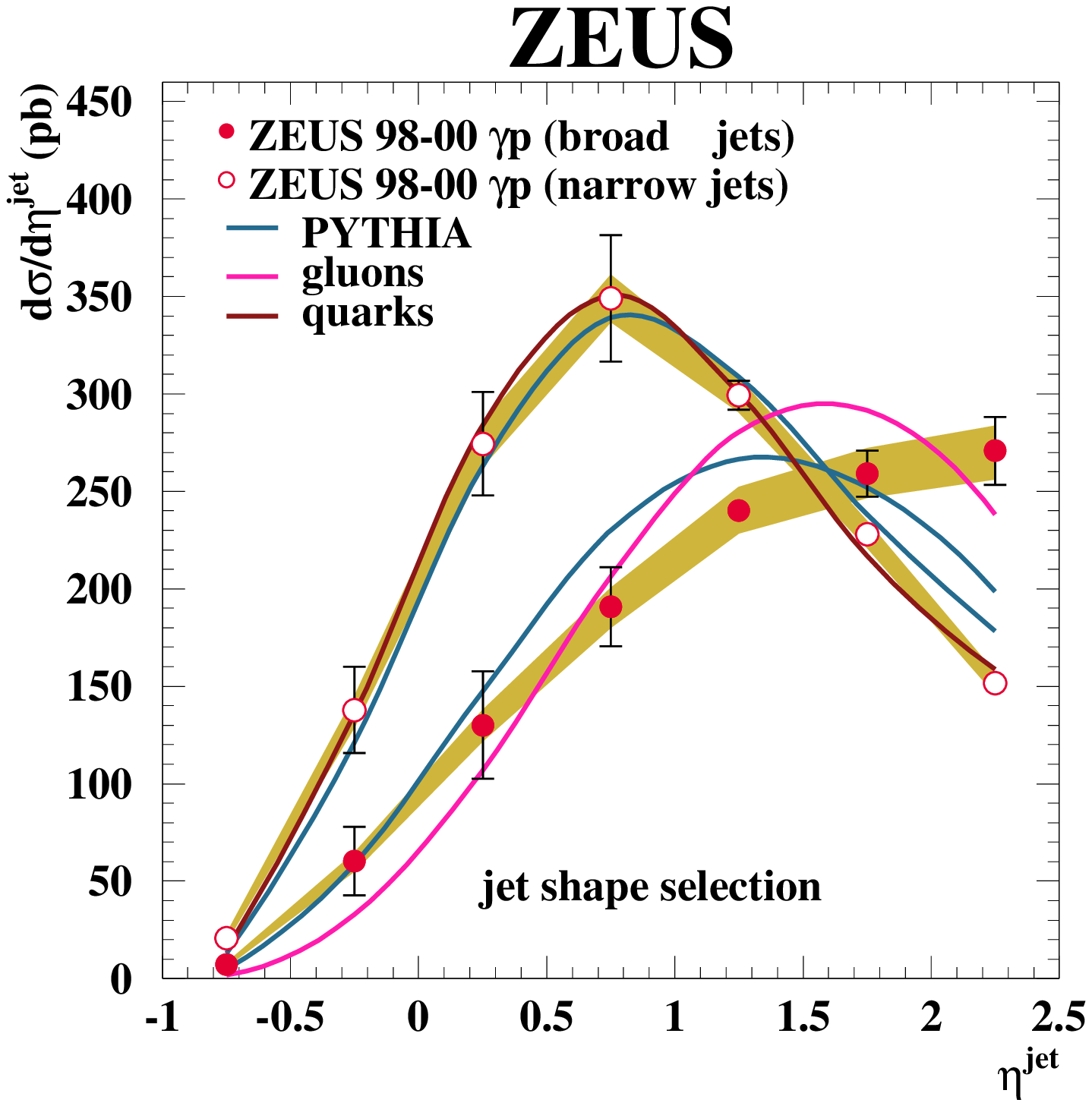,width=10.5cm}}
\put (8.0,-1.0){\epsfig{figure=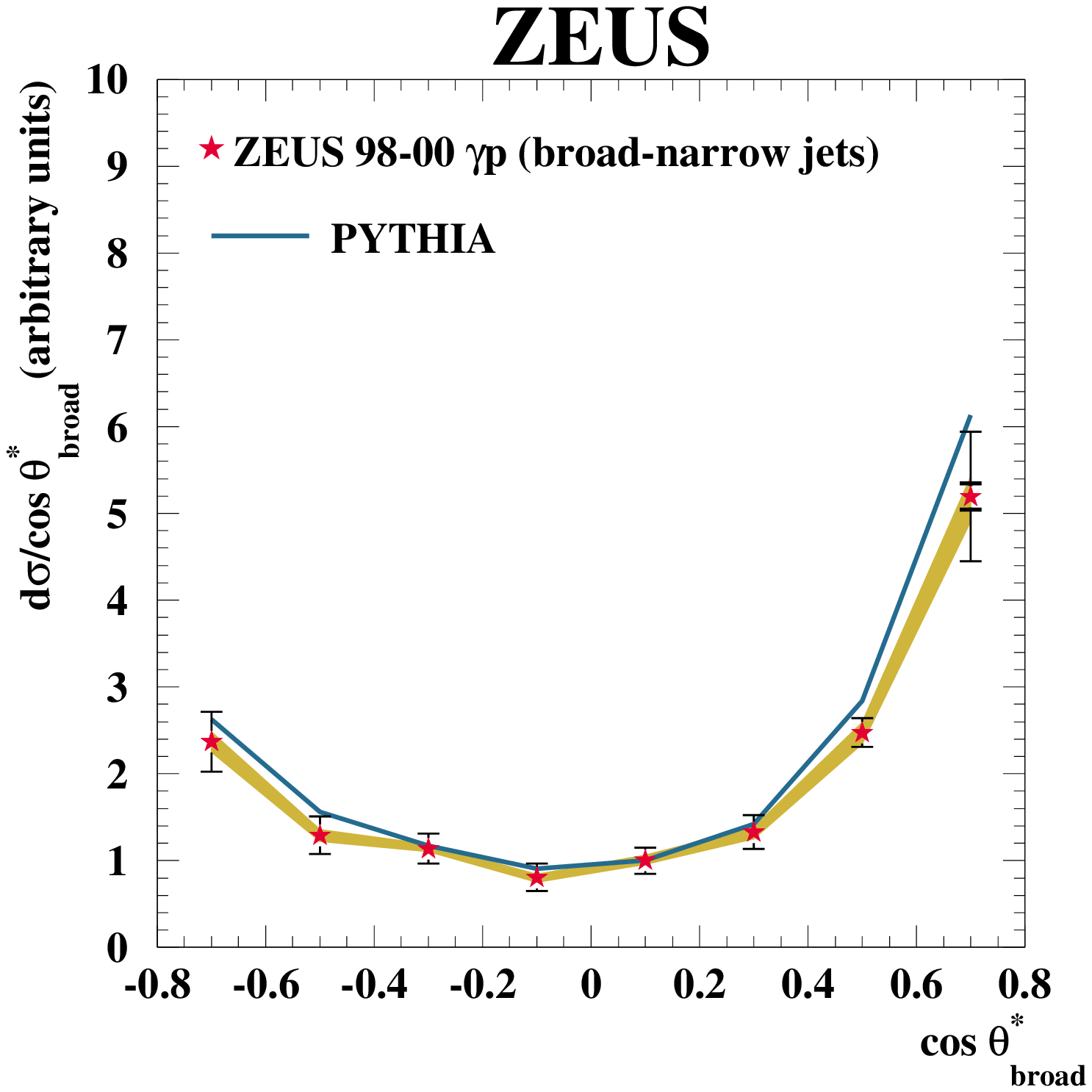,width=10.5cm}}
\put (3.5,-0.2){\bf\small (a)}
\put (12.5,-0.2){\bf\small (b)}
\end{picture}
\caption{(a) Measured inclusive-jet cross section in photoproduction as a
  function of $\etajet$ for samples of broad (dots) and narrow (open
  circles) jets~\protect\cite{inczeus}. (b) Measured dijet cross
  section in photoproduction as a function of $\ccos_{\rm
    broad}$ (stars)~\protect\cite{inczeus}. For comparison, the
  predictions of {\sc Pythia} for jets selected in the same way as in
  the data and for quark- and gluon-initiated jets are included. 
  \label{ten}}
\end{figure}

The differential inclusive-jet cross-section $\seta$ for
photoproduction is shown in Fig.~\ref{ten}a for samples of broad and
narrow jets, separated according to the selection explained above. The
measured cross sections exhibit different behaviours: the $\etajet$
distribution for broad jets increases up to the highest $\etajet$
value measured, whereas the distribution for narrow jets peaks at
$\etajet\approx 0.7$. Monte Carlo calculations using {\sc Pythia}
are compared to the measurements in Fig.~\ref{ten}a. The same
selection method was applied to the jets of hadrons in the MC event
samples and the calculations have been normalised to the total
measured cross section of each sample. The MC predictions provide a
good description of the shape of the narrow-jet distribution in the
data. The shape of the broad-jet distribution in the data is
reasonably well described by {\sc Pythia}. From the calculation of
{\sc Pythia}, the sample of broad jets selected according to the jet
shape is predicted to contain $15\%$ of $gg$ subprocesses in the final
state and $50\%$ of $gq$, and a contamination from processes with only
quarks in the final state of $35\%$. There is a large contribution
from $gq$ final states in the broad-jet sample because the partonic
cross section for the resolved subprocess $q_{\gamma}g_p\rightarrow
qg$ is much larger than the cross section for the subprocesses
$\qq\rightarrow gg$ plus $gg\rightarrow gg$. The sample of narrow jets
contains $62\%$ of $qq$ subprocesses and $34\%$ of $qg$, with a
contamination of $4\%$ from $gg$ subprocesses. Figure~\ref{ten}a also
shows the predictions of {\sc Pythia} for jets of quarks and gluons
separately. These predictions have been obtained without any
selection and are normalised to the data cross sections. The
calculation that includes only quark-initiated jets gives a good
description of the narrow-jet cross section, whereas the calculation
for gluon-initiated jets provides a reasonable description of the
broad-jet cross section. This result supports the expectation that the
broad (narrow)-jet sample is dominated by gluon (quark)-initiated jets.

The distribution in $\theta^*$, where $\theta^*$ is the angle between
the jet-jet axis and the beam direction in the dijet system, reflects
the underlying parton dynamics and is sensitive to the spin of the
exchanged particle. In the case of direct-photon interactions, the
contributing subprocesses at LO QCD involve quark exchange and so
$\scost\propto (1-\cost)^{-1}$ as $\cost\rightarrow 1$. In the case of
resolved-photon interactions, the dominant subprocesses are those that
involve gluon exchange and $\scost\propto (1-\cost)^{-2}$ as
$\cost\rightarrow 1$. The study of the angular distribution for dijet
events with tagged quark- and/or gluon-initiated jets in the final
state, provides then a handle to investigate the underlying parton
dynamics further.

The sample of photoproduced dijet events with one broad jet and
one narrow jet was used to measure $\sccos_{\rm broad}$, where
$\theta^*_{\rm broad}$ refers to the scattering angle measured with
respect to the broad jet. Figure~\ref{ten}b shows the measured dijet
cross section as a function of $\ccos_{\rm broad}$. The measured and
predicted cross sections were normalised to unity at 
$\ccos_{\rm broad}=0.1$. The dijet angular distribution shows a
different behaviour on the negative and positive sides; the measured
cross section at $\ccos_{\rm broad}=0.7$ is approximately twice as
large as at $\ccos_{\rm broad}=-0.7$. The calculation from 
{\sc Pythia} gives a good description of the shape of the measured
$\sccos_{\rm broad}$. The predictions of {\sc Pythia} for the partonic
content are: $52\%$ of $qg$ subprocesses, $4\%$ of $gg$ and $44\%$ of
$qq$. The observed asymmetry is adequately reproduced by the
calculation and is understood in terms of the dominant resolved
subprocess $q_{\gamma}g_p \rightarrow qg$. The $\ccos_{\rm broad}$
distribution for this subprocess is asymmetric due to the different
dominant diagrams in the regions $\ccos_{\rm broad}\rightarrow\pm 1$:
$t$-channel gluon exchange ($\ccos_{\rm broad} \rightarrow +1$) and
$u$-channel quark exchange ($\ccos_{\rm broad} \rightarrow -1$).

In conclusion, the hard subprocesses have been investigated separately
in photoproduction for the first time using the internal structure of
jets.

\section{Conclusions}
\begin{figure}[ht]
\setlength{\unitlength}{1.0cm}
\begin{picture} (10.0,10.5)
\put (5.0,0.0){\epsfig{figure=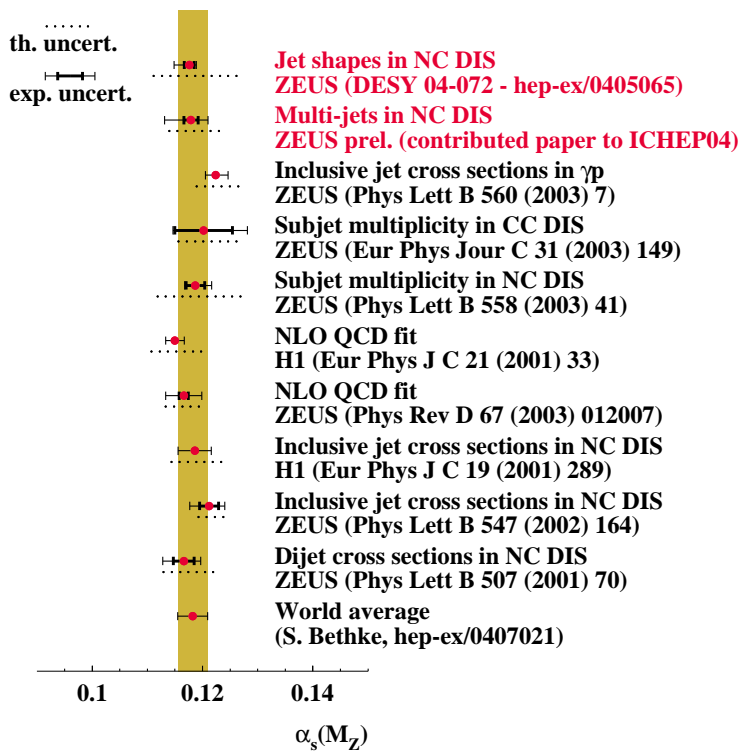,width=11.0cm}}
\end{picture}
\caption{Summary of $\as$ determinations at HERA compared with the
  world average.
  \label{eleven}}
\end{figure}

HERA has become a unique QCD-testing machine. At large scales
considerable progress in understanding and reducing the experimental
and theoretical uncertainties has led to very precise measurements of
the fundamental parameter of the theory, the strong coupling constant
$\as$ (see Fig.~\ref{eleven}) as well as further insight into the
dynamics of quarks and gluons. The use of observables based on the
application of jet algorithms to the hadronic final state of deep
inelastic scattering and of photon-proton interactions leads now to
determinations that are as precise as those coming from more inclusive
measurements, such as from $\tau$ decays. To obtain even better
accuracy in the determination of $\as$, further improvements in the
QCD calculations are needed, e.g. next-to-next-to-leading-order
corrections.

At low $x$ considerable progress has also been obtained in
understanding the mechanisms of parton emission, though the interplay
between the DGLAP, BFKL and CCFM evolution schemes has still to be
fully worked out. Further progress in this respect needs both more
experimental and more theoretical work.

\section*{Acknowledgments}
I would like to thank the organisers for providing a warm atmosphere
conducive to many physics discussions and a well organised
conference. Special thanks to my colleagues from H1 and ZEUS for their
help in preparing this report.

\end{document}